# Dynamic Grouping of Web Users Based on Their Web Access Patterns using ART1 Neural Network Clustering Algorithm

Ramya C.[1], Kavitha G.[2] and Dr. Shreedhara K.S.[3]
[1]M.Tech (Final Year), [2]Lecturer and Professor, [3]Dept. of Studies in CS&E
University B.D.T. College of Engineering, Davangere
Davangere University, Karnataka, India
e-mail: cramyac@gmail.com

*Abstract*—In this paper, we propose ART1 neural network clustering algorithm to group users according to their Web access patterns. We compare the quality of clustering of our ART1 based clustering technique with that of the K-Means and SOM clustering algorithms in terms of inter-cluster and intra-cluster distances. The results show the average inter-cluster distance of ART1 is high compared to K-Means and SOM when there are fewer clusters. As the number of clusters increases, average inter-cluster distance of ART1 is low compared to K-Means and SOM which indicates the high quality of clusters formed by our approach.

## I. INTRODUCTION

We present an ART1 based clustering algorithm to group users according to their Web access patterns. The ART1 is a modified version of ART2 for clustering binary vectors. The advantage of using the ART1 algorithm to group users is that it adapts to the change in users' Web access patterns over time without losing information about their previous Web access patterns. In our ART1 based clustering approach, each cluster of users is represented by a prototype vector that is a generalized representation of URLs frequently accessed by all the members of that cluster. One can control the degree of similarity between the members of each cluster by changing the value of the vigilance parameter. In our work, we analyze the clusters formed by using the ART1 technique by varying the vigilance parameter $\rho$ between the values 0.3 and 0.5. We compare the performance of ART1 clustering technique with that of K-Means and SOM clustering algorithm in terms of average inter-cluster and average intra-cluster distances. Experimental results are provided to show that ART1 NN based clustering approach performs better in terms of intra-cluster and inter-cluster distances compared to K-Means and SOM clustering algorithms. The time complexity of all the three algorithms are compared. ART1 takes less time compared to K-Means and SOM, proving ART1 is efficient than SOM and K-Means.

## II. RELATED WORK

Clustering users based on their Web access patterns is an active area of research in Web usage mining. R. Cooley et al. [1] propose a taxonomy of Web Mining and present various research issues, techniques and future directions in this field. Phoha et al. use competitive neural networks and data mining techniques to develop schemes for fast allocation of Web pages [2]. M. N. Garofalakis et al. [3] review popular data mining techniques and algorithms for discovering Web, hypertext, and hyperlink structure. Y. Fu et al. [4] present a generalization based clustering approach, which combines attribute oriented induction, and BIRCH [5] to generate hierarchical clustering of Web users based on their access patterns. I. Cadez et al. [6] use first-order Markov models to cluster users according to the order in which they request Web pages. The Expectation Maximization algorithm is then used to learn the mixture of first-order Markov models that represent each cluster. G. Paliouras et al. [7] analyze the performance of three clustering algorithms (1) Autoclass, (2) Self Organizing Maps and (3) Cluster Mining for constructing community models for the users of large Websites. Xie and Phoha [8] apply the concept of mass distribution in Dempster-Shafer's theory and propose a belief function similarity measure, which adds to the clustering algorithm the ability to handle uncertainty among Web users' navigation behavior.

## III. CLUSTERING METHODOLOGY

Majority of the techniques that have been used for pattern discovery from Web usage data are clustering methods. A clustering algorithm takes as input a set of input vectors and gives as output a set of clusters thus mapping of each input vector to a cluster. A novel based approach for dynamically grouping Web users based on their Web access patterns using ART1 NN clustering algorithm is presented in this paper. The proposed ART1 NN clustering methodology with a neat architecture is discussed.

*A. The Clustering Model*

The proposed clustering model involves two stages – Feature Extraction stage and the Clustering Stage. First, the features from the preprocessed log data are extracted and a binary pattern vector *P* is generated. Then, ART1 NN clustering algorithm for creating the clusters in the form of prototype vectors is used. The feature extractor forms an input binary pattern vector *P* that is derived from the base vector *D*. The procedure is given in Fig. 1. It generates the pattern vector which is the input vector for ART1 NN based clustering algorithm.

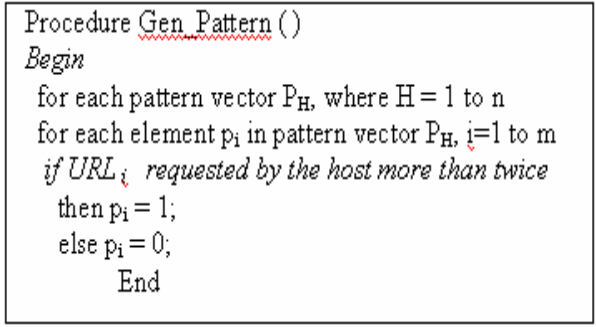

Fig. 1: Procedure for Generating Pattern Vector

*B. ART1 Neural Network Clustering*

Some popular neural networks such as Backpropagation and Self Organizing Maps (SOM) have drawbacks, making them less suitable for solving categorization problems. SOM uses competitive learning and not adaptive in nature. ART1 clustering algorithm is an unsupervised learning algorithm. It is unsupervised in the sense that it establishes the clusters without external interference. It has an advantage that it adapts to changes in user's Web access patterns over time without losing information about their previous Web access patterns. A prototype vector represents each cluster by generalizing the URLs most frequently accessed by all cluster members. Architecture of the ART1 NN based clustering technique for clustering hosts/Web users is as shown in Fig. 2.

The architecture of ART1 NN based clustering is given in Fig. 2. Each input vector activates a winner node in the layer F2 that has highest value among the product of input vector and the bottom-up weight vector. The F2 layer then reads out the top-down expectation of the winning node to F1, where the expectation is normalized over the input pattern vector and compared with the vigilance parameter ρ. If the winner and input vector match within the tolerance allowed by the ρ, the ART1 algorithm sets the control gain G2 to 0 and updates the top-down weights corresponding to the winner. If a mismatch occurs, the gain controls G1 & G2 are set to 1 to disable the current node and process the input on another uncommitted node. Once the network is stabilized, the top-down weights corresponding to each node in F2 layer represent the prototype vector for that node. Summary of the steps involved in ART1 clustering algorithm is shown in Table I.

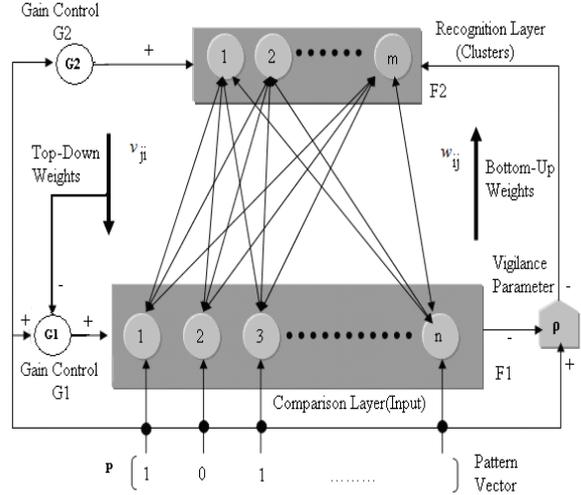

Fig. 2: Architecture of ART1 NN Based Clustering

IV. EXPERIMENTAL RESULTS

The results show that the proposed ART1 algorithm learns relatively stable quality clusters compared to K-Means and SOM clustering algorithms. Here, the quality measures considered are functions of average *Inter-Cluster* and the *Intra-Cluster* distances. Also used are the internal evaluation functions such as *Cluster Compactness* (*Cmp*), *Cluster Separation* (*Sep*) and the combined measure of *Overall Cluster Quality* (*Ocq*) to evaluate the *Intra-Cluster* homogeneity and the *Inter-Cluster* separation of the clustering results. Experimental simulations are also performed using MATLab. Both K-Means and SOM clustering algorithms clusters *N* data points into *k* disjoint subsets $S_j$. The geometric centroid of the data points represents the prototype vector for each subset. SOM is a close cousin of K-Means that embeds the clusters in a low dimensional space right from the beginning and proceeds in a way that places related clusters close together in that space.

Fig. 3 shows the variations in the average inter-cluster distances for the three ART1, K-Means and SOM algorithms. It is observed that, average inter-cluster distance of ART1 is high compared to K-Means and SOM when there are fewer clusters, and as the number of clusters increases, average inter-cluster distance of ART1 is low compared to K-Means and SOM. Variations in the average intra-cluster distances of the three algorithms for varying number of clusters are shown in the Fig. 4. Average intra-cluster distance of ART1 is low compared to K-Means and SOM when there are fewer clusters, and as number of clusters increases, average intra-cluster distance of ART1 is high compared to K-Means and SOM. It is clear from the observation that, the ART1- clustering results are promising compared to K-Means and SOM algorithms.

For the efficiency analysis, the time complexity of all the three algorithms are compared with the same number of hosts to be clustered (input) and with same number of clusters (*k* value in case of SOM and K-Means, $\rho$ value in case of ART1). For a $\rho$ value of 0.5, the response time is measured for the three algorithms on different number of hosts (100,250,500, and 1000) as presented in the Table II. It is observed from the Fig. 5 that, for large data set, ART1 takes less time compared to K-Means and SOM, proving ART1 is efficient than SOM and K-Means. The time complexity of ART1 is almost *linear log time O (n\*log$_2$n),* where as the time complexity of K-Means is *quad log time O (n\*k\*log$_2$n)* and SOM is *polynomial log time O (n\*k\*log$_2$n)* with varying number of iterations.

Fig. 6 shows the GUI of our toolbox with preprocessor tab and ART Clustering tab. The ART Clustering tab allows the user to perform clustering operation. The pattern vector (corresponding to the preprocessed log file) and vigilance parameter values have to be specified as inputs. The prototype vectors of the clusters are displayed at the result panel. The results show that the proposed ART1 algorithm learns relatively stable quality clusters.

V. CONCLUSION

we presented our approach to group hosts (each host represents an organizationally related group of users) according to their Web request patterns. We use the ART1 clustering algorithm to cluster these communities of users. We compare the performance of the ART1 clustering with K-Means and SOM clustering algorithm and show that the ART1 clustering performs better than the K-Means and SOM clustering algorithms. The time complexity of all the three algorithms are compared ART1 takes less time compared to K-Means and SOM, proving ART1 is efficient than SOM and K-Means.

TABLE I: ART1 NN CLUSTERING ALGORITHM

1. Initialize the vigilance parameter $\rho$, $0 \le \rho \le 1$, $w = 2/(1+n)$, $v=1$ where $w$ is $m \times n$ matrix (bottom-up weights) and $v$ is the $n \times m$ matrix (top-down weights), for *n*-tuple input vector and *m* clusters.
2. Binary unipolar input vector $p$ is presented at input nodes. $p_i=\{0,1\}$ for i=1,2,..,n
3. Compute matching scores

$$y_k^0 = \sum_{i=1}^{n} w_{ik} p_i \text{ for k=1,2,...m}$$

Select the best matching existing cluster j with $y_j^0 = Max(y_k^0), k = 1,2,..,m$

4. Perform similarity test for the winning neuron

$$\frac{\sum_{i=1}^{n} v_{ij} p_i}{\|p\|_1} > \rho \quad \text{Where } \rho, \text{ the vigilance parameter and the norm } \|p\|_1 \text{ is the } L_1 \text{ norm defined as,}$$

$\|p\|_1 = \sum_{i=1}^{n} |p_i|$, if the test is passed, the algorithm goes to step 5. If the test fails, then the algorithm goes to step 6, only if top layer has more than a single active node left otherwise, the algorithm goes to step 5.
5. Update the weight matrices for index *j* passing the test (1). The updates are only for entries(*i,j*) where i=1,2,..,m and are computed as follows : $w_{ij}(t+1) = \dfrac{v_{ij}(t) p_i}{0.5 + \sum_{j=1}^{n} v_{ij}(t) p_i}$ and $v_{ij}(t+1) = p_i v_{ij}(t)$

This updates the weights of j$^{th}$ cluster (newly created or the existing one). Algorithm returns to step 2.
6. The node *j* is deactivated by setting $y_j$ to 0. Thus this node does not participate in the current cluster search. The algorithm goes back to step 3 and it will attempt to establish a new cluster different than *j* for the pattern under test.

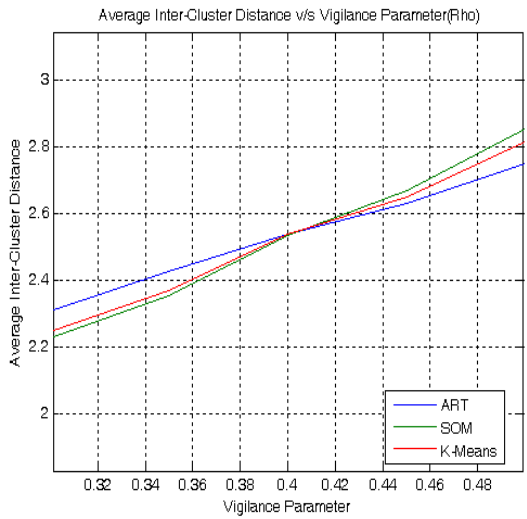

(a) #Hosts: 500

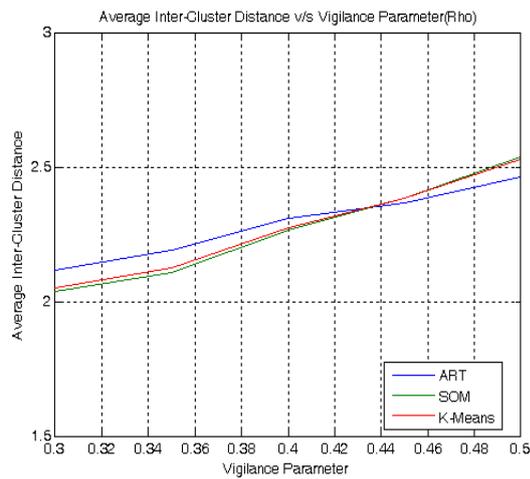

(b) #Hosts: 500

Fig. 3: Variations in Average Inter-Cluster Distances

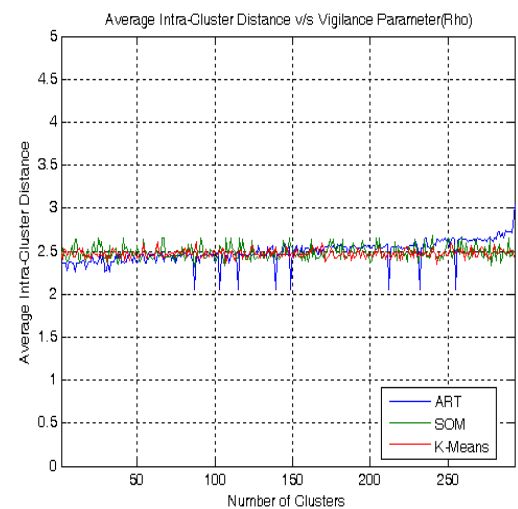

(a) #Hosts: 1000

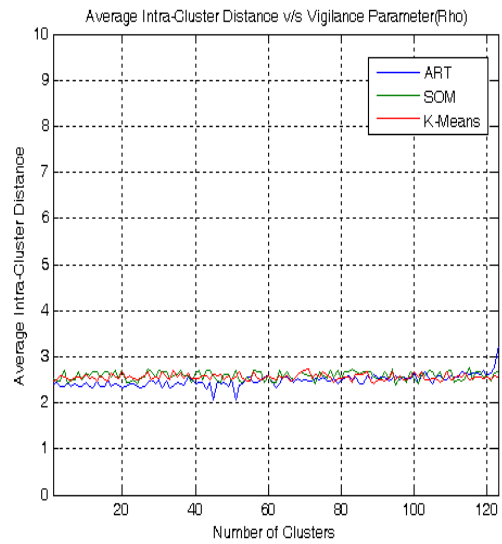

(b) #Hosts: 500

Fig. 4: Variations in Average Intra-Cluster Distances

TABLE II: TIME COMPLEXITY OF CLUSTERING ALGORITHMS

| Hosts | Time (seconds) | | |
|---|---|---|---|
| | ART | K-Means | SOM |
| 100 | 0.156 | 0.188 | 0.422 |
| 250 | 0.875 | 0.931 | 1.797 |
| 500 | 5.156 | 10.219 | 23.359 |
| 1000 | 10.797 | 28.891 | 45.625 |

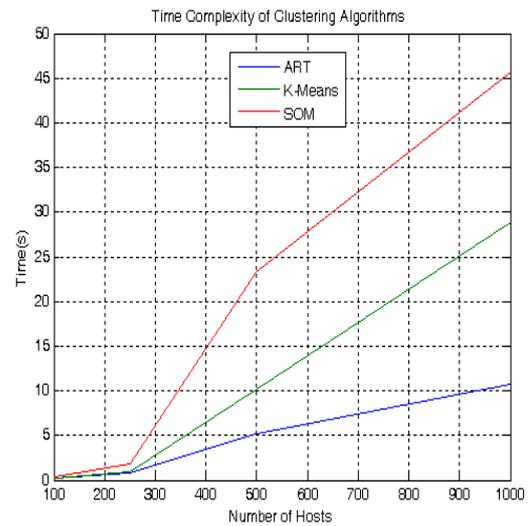

Fig. 5: Time Complexity of Clustering Algorithms

ACKNOWLEDGMENT

Ramya C thanks her brother-in-law Mr. Vivekananda Murthy B G for the interesting suggestions. She also expresses her gratefulness to Dr. Raju G T for guiding this work.

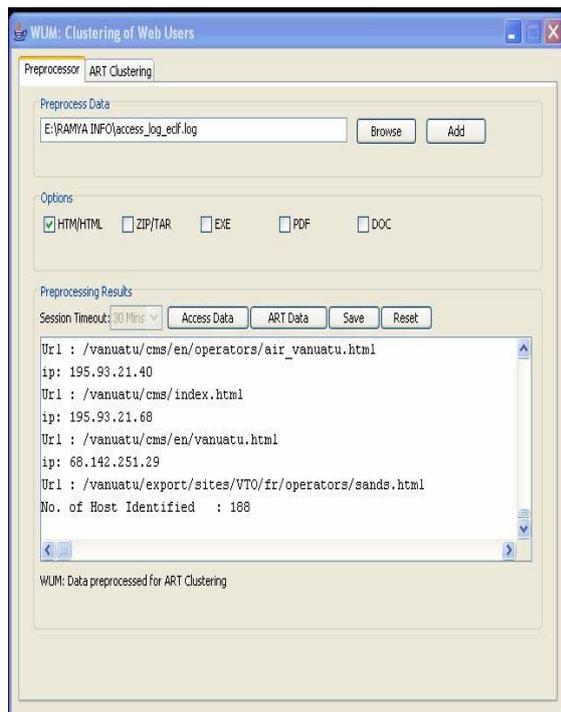

Fig. 6: WUM Toolbox: Results after ART1 Clustering